\documentclass[journal,draftcls,onecolumn,12pt,twoside]{IEEEtran}
\normalsize

\usepackage{lineno}

\usepackage{cite}

\ifCLASSINFOpdf
\else
\fi

\usepackage{amsmath,amsfonts}
\usepackage{algorithmic}
\usepackage{algorithm}
\usepackage{array}
\usepackage[caption=false,font=normalsize,labelfont=sf,textfont=sf]{subfig}
\usepackage{textcomp}
\usepackage{stfloats}
\usepackage{url}
\usepackage{verbatim}
\usepackage{graphicx}
\usepackage{cite}
\usepackage{xcolor}
\usepackage{multirow}

\newcommand{\Ls}{\mathbf{\mathcal{L}}}
\newtheorem{lemma}{Lemma}

\newenvironment{proof}{\par{\it{Proof:}}\ }{\hfill $\square$\par}

\usepackage{algorithm}
\usepackage{algorithmic}
\usepackage{float}
\usepackage{lipsum}
\usepackage{xcolor}
\DeclareMathOperator{\FFT}{FFT}
\DeclareMathOperator{\IFFT}{IFFT}



\newtheorem{aq}{Author Query/Comment}

\newcommand{\baq}{\begin{aq}}
\newcommand{\eaq}{\end{aq}}

\begin{document}
%
\title{A Fast Decoding Algorithm for Generalized Reed-Solomon Codes and Alternant Codes}
%
%
%

\author{Nianqi Tang, Yunghsiang S. Han, \IEEEmembership{Fellow,~IEEE}, Danyang Pei, and Chao Chen 
	\thanks{
		
		Nianqi Tang (724973040@qq.com),  Yunghsiang S. Han (yunghsiangh@gmail.com), and Danyang Pei (202322280103@std.uestc.edu.cn) are with the Shenzhen Institute for Advanced Study, University of Electronic Science and Technology of China, Shenzhen, China. Chao Chen (cchen@xidian.edu.cn) is with the State Key Lab of ISN, Xidian University, Xi'an, China.}
}

\maketitle
\begin{abstract}
In this paper, it is shown that the syndromes of generalized Reed-Solomon (GRS) codes and alternant codes can be characterized in terms of inverse fast Fourier transform, regardless of code definitions. Then a fast decoding algorithm is proposed, which has a computational complexity of $O(n\log(n-k) + (n-k)\log^2(n-k))$  for all $(n,k)$ GRS codes and $(n,k)$ alternant codes. Particularly, this provides a new decoding method for Goppa codes, which is an important subclass of alternant codes. When decoding the binary Goppa code with length $8192$ and correction capability $128$, the new algorithm is nearly 10 times faster than traditional methods. The decoding algorithm is suitable for the McEliece cryptosystem, which is a candidate for post-quantum cryptography techniques.
\end{abstract}
\begin{IEEEkeywords}
generalized Reed-Solomon Codes, alternant codes, decoding algorithms, Goppa codes, fast Fourier transform, McEliece cryptosystem 
\end{IEEEkeywords}

%
\IEEEpeerreviewmaketitle

\section{Introduction}
Generalized Reed-Solomon (GRS) Codes are an important class of error-correcting codes for both theoretical and practical aspects. On the theoretical aspect, GRS codes are maximum-distance-separable (MDS) and are algebraic geometry codes. Furthermore, some alternant codes, which are a subclass of GRS codes, are able to achieve the Gilbert-Varshamov bound. On the practical aspect, several subclasses of GRS codes have been adopted in many applications. For example, Reed-Solomon (RS) codes and Bose-Ray-Chaudhuri-Hocquenghem (BCH) codes have been adopt as error-correcting codes in wired communication systems and storage systems. Goppa codes, a special class of alternant codes, have been adopted in the McEliece cryptosystem, which is chosen as a candidate for post-quantum cryptography (PQC). Hence, investigating the decoding algorithm for GRS codes is of great importance. In the past years, several decoding techniques have been proposed for various subclasses of GRS codes. For example, based on the conventionally defined syndrome, decoding methods for RS codes and BCH codes, which use the Berlekamp-Massey algorithm and the Euclidean algorithm, can be found in \cite{berlekamp2015algebraic} and \cite{sugiyama1975method}, respectively. Based on the remainder polynomial, decoding techniques for RS codes using the Welch-Berlekamp algorithm and the modular approach appeared in \cite{Error1984} and \cite{Fast1995}. A decoding algorithm for Goppa codes was proposed in \cite{patterson1975algebraic} as well as a simple improvement for irreducible Goppa codes. In addition to the hard-decision decoding mentioned above, many soft-decision decoding algorithms were investigated, e.g., Guruswami-Sudan algorithm for RS codes \cite{Guruswami1999improved}, efficient list decoding for RS or BCH codes \cite{wu2012fast, shany2022a}, and so on.

As GRS codes are defined by the finite field Fourier transform, much effort has been devoted to investigating decoding algorithms based on fast Fourier transform (FFT) over finite fields, see \cite{wu2012reduced, bellini2011structure, gao2010additive} for example. There are several challenges when applying an FFT algorithm to design a decoding algorithm. Firstly, whether the obtained decoding algorithms are both available and efficient for various code parameters. Secondly, whether the decoding algorithm has a regular structure such that it is suitable for hardware implementations. In our previous work \cite{tang2022a}, a decoding technique based on Lin-Chung-Han FFT (LCH-FFT, see Section~\ref{sec:preliminary} for details) was proposed for RS codes,  which has the lowest computational complexity to date and has a regular structure. However, the RS codes, which are taken into account in \cite{tang2022a}, are defined in a different way from conventional GRS codes. More precisely, the $(n,k)$ RS code in \cite{tang2022a} is defined by
\begin{align*}
	\{(f(\omega_0), f(\omega_1), \dots, f(\omega_{n-1}))\mid \deg(f(x)) < 2^m - n + k,\\ f(\omega_j) = 0, j = n, n + 1, \dots, 2^m - 1 \},
\end{align*}
where $f(x)\in GF(2^m)[x]$ and $\omega_j$, whose definition can be found in Section \ref{sec:LCH}, are distinct elements in finite field $GF(2^m)$. This $(n,k)$ RS code can be seen as deleting message symbols for a $(2^m, 2^m - n + k)$ RS code. However, a conventional GRS code is defined as
\begin{equation*}
	\{w_0f(\alpha_0), w_1f(\alpha_1), \dots, w_{n-1}f(\alpha_{n-1})\mid \deg(f(x)) < k\},
\end{equation*}
where $w_i$ are nonzero elements and $\alpha_i$ are distinct elements in $GF(2^m)$. The conventional $(n,k)$ GRS code can be seen as deleting parity symbols from a $(2^m, k)$ GRS code. Hence, it is not straightforward whether a fast decoding algorithm based on FFT exists for arbitrary GRS codes.
On the other hand, it is not clear whether such a fast decoding algorithm exists for alternate codes, such as Goppa codes, since their definitions are not related to the Fourier transform. This paper is devoted to addressing these issues.

In this paper, we first investigate the concept of generalized syndromes. The concept of generalized syndromes was first shown in \cite{araki1992generalized}, which was used for illustrating the relationship between the conventional decoding key equation and Welch-Berlekamp's key equation. We prove that, by specifying the polynomial $T(x)$ (which shall be defined shortly), the generalized syndrome can be characterized in terms of the high degree part of the inverse FFT  (IFFT) of the received vector. Accordingly, an efficient method for computing this type of syndrome is also derived, which is of complexity $O(n\log(n - k))$ for code length $n$ and information dimension $k$. Notably, this type of syndromes is defined with respect to the parity check matrix, regardless of code definitions. As $T(x)$ is fixed, all GRS codes and alternant codes are shown to have the same type of key equation. This leads to a unified decoding algorithm with the computational complexity of $O(n\log(n-k) + (n-k)\log^2(n-k))$, which is the lowest computational complexity to date according to the best of our knowledge. As an example, we investigate the proposed algorithm regarding the Goppa codes. A computational complexity comparison among the proposed algorithm, the famous Patterson method in \cite{patterson1975algebraic}, and a conventional decoding method shown in \cite{macwilliams1977theory} is provided. When decoding the Goppa code of length $8192$, which is defined on $GF(2^{13})$ and has Goppa polynomial $x^{128} + x^7 + x^2 + x + 1$ (this Goppa code has been submitted to National Institute of Standards and Technology (NIST) as a candidate for post-quantum cryptography, see \cite{NIST_PQC}), the proposed algorithm is nearly 10 times faster than other methods. This shows that the proposed algorithm is superior in terms of computational complexity for practical use.

The main contributions of this paper can be summarized as follows:
\begin{enumerate}
	\item We characterize the generalized syndrome in terms of the IFFT for all GRS codes and alternant codes and provide an efficient method for computing the syndrome;
	\item Based on the above syndrome, we show that all GRS codes and alternant codes have the same type of key equation. This leads to a fast decoding algorithm with complexity $O(n\log(n-k) + (n-k)\log^2(n-k))$ for any $(n,k)$ GRS code and alternant code, which is the best complexity to date. This also provides another derivation of the key equation in \cite{Lin20161}, which is more straightforward and comprehensive;
	\item Based on the algorithm, we provide a fast decoding algorithm for Goppa codes and show that the algorithm is efficient for practical use.
\end{enumerate}

The paper is organized as follows. Section \ref{sec:preliminary} first reviews the definition of GRS codes and alternant codes, and a brief introduction to LCH-FFT is then provided. Section \ref{sec:syndrome} discusses the generalized syndrome and characterizes the generalized syndrome in terms of the IFFT of the received vector. Section \ref{sec:dec} proposes a decoding algorithm for any GRS code and alternant code. Section \ref{sec:dec} provides the application of the proposed method to separable Goppa codes and gives a complexity comparison between the proposed algorithm and traditional methods when decoding Goppa codes. Finally, Section \ref{sec:conclusion} concludes the paper and presents several open issues.

\section{Preliminary}\label{sec:preliminary}
This section reviews some basic definitions and concepts, including the GRS codes and alternant codes, and provides a brief introduction to the FFT over a finite field (Lin-Chung-Han FFT), which is crucial for designing a decoding algorithm. Throughout this paper, we mainly focus on codes defined over binary extension field $GF(2^m)$ since they are of great importance in practice. Nevertheless, the technologies presented here can be generalized to any finite field.
\subsection{Generalized Reed-Solomon Codes and Alternant Codes}

Let $n$ and $k$ be integers satisfying $0 < k \leq n\leq 2^m$. Let $\Ls = (\alpha_0, \alpha_1, \dots, \alpha_{n - 1})$ be $n$ distinct elements in $GF(2^m)$ and let $\mathbf{w} = (w_0, w_1, \dots, w_{n - 1})$ whose elements are all nonzero in $GF(2^m)$. The generalized Reed-Solomon (GRS) code $GRS_k(\Ls, \mathbf{w})$ consists of all vectors
\begin{equation}
	(w_0 f(\alpha_0), w_1 f(\alpha_1), \dots, w_{n - 1} f(\alpha_{n - 1})),
\end{equation}
where $f(x)\in GF(2^m)[x]$ satisfying $\deg(f(x)) < k$. $GRS_k(\Ls, \mathbf{w})$ code is an $(n, k, d)$ code over $GF(2^m)$, where $d = n - k + 1$ is the minimum distance.

The dual code of $GRS_k(\Ls, \mathbf{w})$ is also a GRS code. The parity check matrix of $GRS_k(\Ls, \mathbf{w})$ can be written as 
\begin{align*}
	H &= \left(
	\begin{array}{cccc}
		y_0 & y_1 & \dots & y_{n - 1} \\
		y_0\alpha_0 & y_1\alpha_1 & \dots & y_{n - 1}\alpha_{n - 1}\\
		\vdots & \vdots & \dots & \vdots\\
		y_0\alpha_0^{n - k - 1} & y_1\alpha_1^{n - k - 1} & \dots & y_{n - 1}\alpha_{n - 1}^{n - k - 1}
	\end{array}
	\right),
\end{align*}
where $\mathbf{y} = (y_0, y_1, \dots, y_{n - 1})\in GF(2^m)$ is a vector whose components are all nonzero. It can be shown that
\begin{equation}\label{eq:inverse}y_i=w_i^{-1}\left(\prod_{0\le j\le n-1,j\neq i}(\alpha_i-\alpha_j)\right)^{-1}.\end{equation}

The alternant code $\mathcal{A}_k(\Ls, \mathbf{y})$ consists of all codewords of $GRS_k(\Ls, \mathbf{w})$ whose components all lie in $GF(2)$. In other words, $\mathcal{A}_k(\Ls, \mathbf{y})$ is the subfield subcode of $GRS_k(\Ls, \mathbf{w})$. $\mathcal{A}_k(\Ls, \mathbf{y})$ is an $(n, \geq n - m(n - k), \geq d)$ code over $GF(2)$, where $d = n - k + 1$. Straightforwardly, $H$ is also the parity check matrix of $\mathcal{A}_k(\Ls, \mathbf{y})$.

In subsequent sections, unless other specifications, we assume that the codes that appear are determined by the parity check matrix $H$. We do not distinguish $GRS_k(\Ls, \mathbf{w})$ and $\mathcal{A}_k(\Ls, \mathbf{y})$ since the decoding algorithm focuses mainly on $H$.

\subsection{Lin-Chung-Han FFT (LCH-FFT)\cite{4319038}}~\label{sec:LCH}


Let $\{v_0, v_1, \dots, v_{m-1}\}$ be a basis of $GF(2^m)$ over $GF(2)$. The elements of $GF(2^m)$ can be represented by
\begin{equation*}
	\omega_j = j_0v_0 + j_1v_1 + \dots + j_{m-1}v_{m-1}, 0\leq j< 2^m - 1,
\end{equation*}
where $(j_0, j_1, \dots, j_{m-1})$ is the binary representation of the integer $j$. The subspace polynomial $s_{\tau}(x)$ is defined as
\begin{equation*}
	s_{\tau}(x) = \prod_{j = 0}^{2^{\tau} - 1}(x - \omega_j), 0\leq \tau\leq m.
\end{equation*}
Define the polynomial ${X}_j(x)$ as
\begin{equation*}
	{X}_j(x) = s_0(x)^{j_0}s_1(x)^{j_1}\cdots s_{m-1}(x)^{j_{m-1}}.
\end{equation*}
Then the set ${\mathbb{X}} = \{{X}_0(x), {X}_1(x), \dots, {X}_{2^m - 1}(x)\}$ is a basis of $GF({2^m})[x] / (x^{2^m} - x)$ over $GF({2^m})$ since $\deg({X}_j(x)) = j$. Let
\begin{equation*}
	\bar{X}_{j}(x) = {X}_j(x) / p_j, 0 \leq j < 2^m - 1,
\end{equation*}
where $p_j = s_0(v_0)^{j_0}s_1(v_1)^{j_1}\cdots s_{m-1}(v_{m-1})^{j_{m-1}}$. It is easy to check that the set $\bar{\mathbb{X}} = \{\bar{X}_0(x), \bar{X}_1(x), \dots, \bar{X}_{2^m - 1}(x) \}$ is also a basis of $GF({2^m})[x] / (x^{2^m} - x)$ over $GF({2^m})$.

For a polynomial $f(x) \in GF({2^m})[x] / (x^{2^m} - x)$ of degree less than $2^{\tau}$, given its coordinate vector $\bar{\mathbf{f}} = (\bar{f}_0, \bar{f}_1, \dots, \bar{f}_{2^{\tau} - 1})$ with respect to $\bar{\mathbb{X}}$ and $\beta\in GF({2^m})$, the LCH-FFT computes $$\mathbf{F} = (f(\omega_0 + \beta), f(\omega_1 + \beta), \dots, f(\omega_{2^{\tau} - 1} + \beta))$$ within $O(2^{\tau}\log(2^{\tau}))$ field operations, which is denoted by
\begin{equation*}
	\mathbf{F} = \text{FFT}_{\bar{\mathbb{X}}}(\bar{\mathbf{f}}, \tau, \beta).
\end{equation*}
Its inverse transform is written as
\begin{equation*}
	\bar{\mathbf{f}} = \text{IFFT}_{\bar{\mathbb{X}}}(\mathbf{F}, \tau, \beta).
\end{equation*}
Detailed descriptions of $\FFT_{{\bar{\mathbb{X}}}}$ and $\IFFT_{{\bar{\mathbb{X}}}}$ are shown in Algorithms \ref{FFTX} and \ref{IFFTX}, respectively. For more discussions, please refer to \cite{Lin20161}. It should be noted that the LCH-FFT has been generalized to evaluate any number of points. More details can be found in \cite{2207.11079}.

\begin{algorithm}[ht]
	\caption{$\FFT_{{\bar{\mathbb{X}}}}$~\cite{4319038}}
	\label{FFTX}
	\begin{algorithmic}[1]
		\REQUIRE
		$\bar{\mathbf{f}} = (\bar{f}_0,\bar{f}_1,\dots,\bar{f}_{2^\tau-1})$, $\tau$, $\beta$.
		\ENSURE
		$(f(\omega_0 + \beta), f(\omega_1 + \beta), \dots, f(\omega_{2^\tau-1} + \beta))$.
		
		\IF {$\tau = 0$} \RETURN $\bar{f}_0$
		\ENDIF
		\FOR {$l = 0, 1,\dots,2^{\tau-1}-1$}
		\STATE {$a_l^{(0)} = \bar{f}_l + \dfrac{s_{\tau-1}(\beta)}{s_{\tau-1}(v_{\tau-1})}\bar{f}_{l+2^{\tau-1}}$}
		\STATE {$a_l^{(1)} = a_l^{(0)} + \bar{f}_{l+2^{\tau-1}}$}
		\ENDFOR
		\STATE $\mathbf{a}^{(0)}=(a_0^{(0)},\dots,a_{2^{\tau-1}-1}^{(0)}),\mathbf{a}^{(1)}=(a_0^{(1)},\dots,a_{2^{\tau-1}-1}^{(1)})$
		\STATE Calculate $\mathbf{A}_0 = \FFT_{\bar{\mathbb{X}}}(\mathbf{a}^{(0)},\tau-1,\beta)$, $\mathbf{A}_1 = \FFT_{\bar{\mathbb{X}}}(\mathbf{a}^{(1)},\tau-1,v_{\tau-1} + \beta)$
		\RETURN $(\mathbf{A}_0,\mathbf{A}_1)$
	\end{algorithmic}
\end{algorithm}

\begin{algorithm}[ht]
	\caption{$\IFFT_{{\bar{\mathbb{X}}}}$~\cite{4319038}}
	\label{IFFTX}
	\begin{algorithmic}[1]
		\REQUIRE
		$\mathbf{F} = (f(\omega_0 + \beta), f(\omega_1 + \beta), \dots, f(\omega_{2^\tau-1} + \beta)), \tau, \beta$
		\ENSURE
		$\bar{\mathbf{f}}$ such that $ \mathbf{F}= \FFT_{\bar{\mathbb{X}}}(\bar{\mathbf{f}},\tau,\beta)$
		
		\IF {$\tau = 0$}
		\RETURN $f(\omega_0 + \beta)$
		\ENDIF
		
		\STATE  $\mathbf{A}_{0}=(f(\omega_0 + \beta),\dots,f(\omega_{2^{\tau-1}-1} + \beta)),\mathbf{A}_{1}=(f(\omega_{2^{\tau-1}} + \beta)),\dots,f(\omega_{2^{\tau}-1} + \beta))$
		\STATE $\mathbf{a}^{(0)} = \IFFT_{\bar{\mathbb{X}}}(\mathbf{A}_0,\tau-1,\beta)$
		, $\mathbf{a}^{(1)} = \IFFT_{\bar{\mathbb{X}}}(\mathbf{A}_1,\tau-1,v_{\tau-1} + \beta)$
		
		\FOR {$l = 0, 1,\dots,2^{\tau-1}-1$}
		\STATE {$\bar{f}_{l+2^{\tau-1}} = a_l^{(0)} + a_l^{(1)}$}
		\STATE {$\bar{f}_l = a_l^{(0)} + \dfrac{s_{\tau-1}(\beta)}{s_{\tau-1}(v_{\tau-1})}\bar{f}_{l+2^{\tau-1}}$}
		\ENDFOR
		\RETURN $\mathbf{\bar{f}}$
	\end{algorithmic}
\end{algorithm}

\section{Generalized Syndrome}\label{sec:syndrome}

In this section, we first review the definition of generalized syndromes, which was first proposed in \cite{araki1992generalized}. Then, we prove that a specific generalized syndrome can be characterized in terms of the high degree part of the IFFT of the received vector. In the next section, based on this specific generalized syndrome, we shall propose a fast decoding algorithm of complexity $O(n\log(n-k) + (n-k)\log^2(n-k))$ for any GRS and alternant codes.

For a codeword $\mathbf{c} = (c_0, c_1, \dots, c_{n-1})$, the received vector can be written as
\begin{align*}
	\mathbf{r} &= (r_0, r_1, \dots, r_{n - 1})\\
	&= (c_0, c_1, \dots, c_{n-1}) + (e_0, e_1, \dots, e_{n-1})\\
	&= \mathbf{c} + \mathbf{e},
\end{align*}
where $\mathbf{e} = (e_0, e_1, \dots, e_{n-1})$ is the error pattern. If $e_i\neq 0$, an error occurs at position $i$. The error location set $E$ is defined as
$E = \{i|e_i\neq 0, i = 0, 1, \dots, n - 1\}$.

For any polynomial $T(x)\in GF(2^m)[x]$ of degree $n-k$, the generalized syndrome of $GRS_k(\Ls, \mathbf{w})$ and $\mathcal{A}_k(\Ls, \mathbf{y})$ is defined as
\begin{equation}\label{eq:syndrome}
	\mathbf{S}(x) = \sum_{i = 0}^{n - 1}r_iy_i\frac{T(x) - T(\alpha_i)}{x - \alpha_i}.
\end{equation}

\begin{lemma}
	The generalized syndrome satisfies that
	\begin{equation*}
		\mathbf{S}(x) = \sum_{i\in E}e_iy_i\frac{T(x) - T(\alpha_i)}{x - \alpha_i}.
	\end{equation*} 
	\begin{proof}
		This proof can be found in \cite{araki1992generalized}. However, we repeat the proof here to make this paper self-contained.
		
		As $H\mathbf{c}^{T} = 0$, a codeword $\mathbf{c} = (c_0, c_1, \dots, c_{n-1})$ satisfies
		\begin{equation*}
			\sum_{i = 0}^{n - 1}c_iy_i\alpha_i^l = 0,\ \text{for}\ l = 0, 1, \dots, n - k - 1.
		\end{equation*}
		If we write $T(x) = \sum_{j = 0}^{n - k} T_j x^j$, we have
		\begin{align*}
			T(x) - T(\alpha_i) &= \sum_{j = 0}^{n - k}T_j (x^j - \alpha_{i}^j)\\
			&= \sum_{j = 1}^{n - k}T_j (x - \alpha_i)\sum_{l = 0}^{j - 1}x^{j - 1 - l}(\alpha_{i})^l.
		\end{align*}
		This leads to
		\begin{equation*}
			\frac{T(x) - T(\alpha_i)}{x - \alpha_i} = \sum_{j = 1}^{n - k}T_j \sum_{l = 0}^{j - 1}x^{j - 1 - l}\alpha_{i}^l.
		\end{equation*}
		Hence, one has
		\begin{align*}
			\mathbf{S}(x) &=  \sum_{i = 0}^{n - 1}r_iy_i\sum_{j = 1}^{n - k}T_j \sum_{l = 0}^{j - 1}x^{j - 1 - l}\alpha_{i}^l\\
			&= \sum_{j = 1}^{n - k}T_j \sum_{l = 0}^{j - 1}x^{j - 1 - l}\sum_{i = 0}^{n - 1}r_iy_i\alpha_i^l\\
			&= \sum_{j = 1}^{n - k}T_j \sum_{l = 0}^{j - 1}x^{j - 1 - l}\sum_{i = 0}^{n - 1}(c_i + e_i)y_i\alpha_i^l\\
			&= \sum_{j = 1}^{n - k}T_j \sum_{l = 0}^{j - 1}x^{j - 1 - l}\sum_{i\in E}e_iy_i\alpha_i^l\\
			&= \sum_{i\in E}e_iy_i\sum_{j = 1}^{n - k}T_j \sum_{l = 0}^{j - 1} x^{j - 1 - l}\alpha_i^l\\
			&= \sum_{i\in E}e_iy_i\frac{T(x) - T(\alpha_i)}{x - \alpha_i}.
		\end{align*}
	\end{proof}
\end{lemma}

Based on the generalized syndrome defined above, the corresponding decoding procedure is described as follows.
Define the error locator polynomial
\begin{equation*}
	\lambda(x) = \prod_{i\in E}(x - \alpha_i).
\end{equation*}
The key equation is 
\begin{equation}\label{eq:key}
	\mathbf{S}(x)\lambda(x) = q(x)T(x) + z(x),
\end{equation}
where
\begin{align*}
	q(x) &= \sum_{i\in E}e_iy_i\prod_{j\in E \atop j\neq i}(x - \alpha_j),\\
	z(x) &= -\sum_{i\in E}e_iy_iT(\alpha_i)\prod_{j\in E \atop j\neq i}(x - \alpha_j).
\end{align*}
Clearly, $\deg(z(x)) < \deg(\lambda(x))$. Solving the key equation, one can obtain $\lambda(x)$ and $z(x)$. By \eqref{eq:key}, one can obtain $q(x)$. 
The error locations can be computed by finding the roots of $\lambda(x)$, and the error values $e_i$ can be computed by
\begin{equation*}
	e_i = \frac{q(\alpha_i)}{y_i^{-1}\lambda{'}(\alpha_i)},
\end{equation*}
where $\lambda{'}(x)$ denotes the formal derivative of $\lambda(x)$.

The above decoding procedure works for any $T(x)$. In the following, we prove that, by specifying $T(x)$, the generalized syndrome can be related to the IFFT of $\mathbf{r}$. We now specify $T(x) = \prod_{j = 0}^{n - k - 1}(x -\omega_j)$ hereafter, in which the definition of $\omega_j$ appears in Section \ref{sec:LCH}. 

The elements in $GF(2^m)$ can be arranged as 
\begin{equation*}
	(\alpha_0, \alpha_1, \dots, \alpha_{n-1}, \alpha_{n}, \dots, \alpha_{2^m-1}),
\end{equation*}
where $\alpha_0, \alpha_2, \dots, \alpha_{n-1}$ have the same order with $\mathcal{L}$ and the residual elements in $GF(2^m)\setminus\mathcal{L}$ are with arbitrary order. Let $\pi$ denotes a permutation between the indices of $(\omega_0, \omega_1, \dots, \omega_{2^m-1})$ and $(\alpha_0, \alpha_1, \dots, \alpha_{2^m-1})$, in which $\pi(j) = i$ if and only if $\omega_j = \alpha_i$. The inverse mapping of $\pi$ is written by $\pi^{-1}$.

Construct a new vector 
\begin{equation*}
	\mathbf{r}^{'} = (r_0^{'}, r_1^{'}, \dots, r_{2^m-1}^{'}),
\end{equation*}
satisfying $r_j^{'}  = 0$ for all $j$ such that $\pi(j) \geq n$ and satisfying $r^{'}_j = r_{\pi{(j)}}y_{\pi(j)}$ otherwise. 

We are now able to rewrite the generalized syndrome \eqref{eq:syndrome} as
\begin{equation}\label{eq:new-symdrome}
	\mathbf{S}(x) = \sum_{j = 0}^{2^m - 1} r_{j}^{'}\frac{T(x) - T(\omega_j)}{x - \omega_j}.
\end{equation}

For the vector $(r_0^{'}, r_1^{'}, \dots, r_{2^m - 1}^{'})$, there exists an unique polynomial $f(x)\in GF(2^m)[x]$ of degree less than $2^m$ such that 
$$f(\omega_j) = r_j^{'}$$
for all $j$. 
By Lagrange interpolation, $f(x)$ can be represented by
\begin{equation*}
	f(x) = \sum_{j = 0}^{2^m - 1}r_j^{'}\frac{s_m(x)}{x - \omega_j},
\end{equation*}
where $s_m(x) = \prod_{l = 0}^{2^m-1}(x - \omega_l)$. Let $\mu$ be the smallest integer such that $\epsilon = 2^{\mu} \geq n-k$. Define
\begin{equation*}
	\mathbf{S}_1(x) = \sum_{j = 0}^{2^m - 1} r_{j}^{'}\frac{s_{\mu}(x) - s_{\mu}(\omega_j)}{x - \omega_j}.
\end{equation*}

\begin{lemma}\label{lem:s1x}
	$\mathbf{S}_1(x)$ is the quotient of $f(x)$ divided by ${\mathbb{X}}_{2^m - 2^{\mu}}(x)$, i.e.,
	\begin{equation}\label{eq:synd1}
		f(x) = \mathbf{S}_1(x){\mathbb{X}}_{2^m - 2^{\mu}}(x) + \eta_1(x),
	\end{equation}
	where $\deg(\eta_1(x)) < \deg({\mathbb{X}}_{2^m - 2^{\mu}}(x))=2^m - 2^{\mu}$.
	
	\begin{proof}
		The case that $\mu = m$ is trivial since ${\mathbb{X}}_0(x) = 1$ and $s_{m}(\omega_j)=0$ for $0\le j\le 2^m-1$.
		
		If $\mu = 0$, one must have $\mathbf{S}_1(x) = \sum_{j=0}^{2^m-1}r_j'$ and the claim obviously holds.
		
		Now we assume that $0 < \mu < m$. $s_m(x)$ can be written as 
		\begin{equation*}
			s_m(x) = (s_{\mu}(x) + s_{\mu}(v_{\mu})){\mathbb{X}}_{2^m - 2^{\mu}}(x) + \eta_2(x),
		\end{equation*}
		where $\deg(\eta_2(x)) < \deg({\mathbb{X}}_{2^m - 2^{\mu}}(x))$ (This can be referred to \cite[Equation (75)]{FFT2016}). It follows that
		\begin{equation}\label{eq:s_m_divide1}
			\frac{s_m(x)}{x - \omega_j} = \frac{(s_{\mu}(x) + s_{\mu}(v_{\mu})){\mathbb{X}}_{2^m - 2^{\mu}}(x) + \eta_2(x)}{x - \omega_j}.
		\end{equation}
		
		Given the polynomial $\frac{s_{\mu}(x) - s_{\mu}(\omega_j)}{x - \omega_j}{\mathbb{X}}_{2^m - 2^{\mu}}(x)$, we have
		\begin{align}
			\frac{s_m(x)}{x - \omega_j} 
			&= \frac{s_{\mu}(x) - s_{\mu}(\omega_j)}{x - \omega_j}{\mathbb{X}}_{2^m - 2^{\mu}}(x) + \eta_3(x)\label{eq:new_divide}\\
			&= \frac{(s_{\mu}(x) - s_{\mu}(\omega_j)){\mathbb{X}}_{2^m - 2^{\mu}}(x) + \eta_3(x)(x - \omega_j)}{x - \omega_j}.\label{eq:s_m_divide2}
		\end{align}
		Combining  \eqref{eq:s_m_divide1} and \eqref{eq:s_m_divide2}, this leads to
		\begin{equation*}
			(s_{\mu}(v_{\mu}) + s_{\mu}(\omega_j)){\mathbb{X}}_{2^m - 2^{\mu}}(x) + \eta_2(x) - \eta_3(x)(x - \omega_j) = 0.
		\end{equation*}
		If $s_{\mu}(v_{\mu}) + s_{\mu}(\omega_j) = 0$, we must have $\deg(\eta_3(x)) < \deg(\eta_2(x)) < \deg(\mathbb{X}_{2^m - 2^{\mu}}(x))$. On the other hand, if $s_{\mu}(v_{\mu}) + s_{\mu}(\omega_j) \neq 0$, it is obviously that $\deg(\eta_3(x)) < \deg(\mathbb{X}_{2^m - 2^{\mu}}(x))$. Hence, according to \eqref{eq:new_divide}, we can conclude that $\frac{s_{\mu}(x) - s_{\mu}(\omega_j)}{x - \omega_j}$ is the quotient of $\frac{s_m(x)}{x - \omega_j}$ divided by $\mathbb{X}_{2^m - 2^{\mu}}(x)$. Finally, by summation on $j$, it is straightforward to see that $\mathbf{S}_1(x)$ is the quotient of $f(x)$ divided by ${\mathbb{X}}_{2^m - 2^{\mu}}(x)$ and the claim follows.
	\end{proof}
\end{lemma}

\begin{lemma}\label{lem:sx}
	$\mathbf{S}(x)$ is the quotient of $\mathbf{S}_1(x)$ divided by $\prod_{l = n - k}^{\epsilon - 1}(x - \omega_l)$, i.e.,
	\begin{equation}\label{eq:syndrome2}
		\mathbf{S}_1(x) = \mathbf{S}(x)\prod_{l = n - k}^{\epsilon - 1}(x - \omega_l) +\eta_4(x),
	\end{equation}
	where $\deg(\eta_4(x)) < \deg(\prod_{l = n - k}^{\epsilon - 1}(x - \omega_l))$.
	\begin{proof}
		If $n - k = \epsilon$, we have $\mathbf{S}(x) = \mathbf{S}_1(x)$ and $\prod_{l = n - k}^{\epsilon - 1}(x - \omega_l) = 1$. The claim obviously holds.
		
		If $n - k < \epsilon$, we have
		\begin{equation*}
			s_{\mu}(x) = T(x)\prod_{l = n - k}^{\epsilon - 1}(x - \omega_l).
		\end{equation*}
		So
		\begin{equation}\label{eq:s_mu_divide1}
			\frac{s_{\mu}(x) - s_{\mu(\omega_j)}}{x - \omega_j} = \frac{T(x)\prod_{l = n - k}^{\epsilon - 1}(x - \omega_l) - s_{\mu(\omega_j)}}{x - \omega_j}.
		\end{equation}
		Given the polynomial $\frac{T(x) - T(\omega_j)}{x - \omega_j}\prod_{l = n - k}^{\epsilon - 1}(x - \omega_l)$, one can write
		\begin{align}
			&\frac{s_{\mu(x)} - s_{\mu(\omega_j)}}{x - \omega_j} \notag\\=
			&\frac{T(x) - T(\omega_j)}{x - \omega_j}\prod_{l = n - k}^{\epsilon - 1}(x - \omega_l) + \eta_5(x)\label{eq:new_divide1}\\
			=&\frac{T(x)\prod_{l = n - k}^{\epsilon - 1}(x - \omega_l) - T(\omega_j)\prod_{l = n - k}^{\epsilon - 1}(x - \omega_l)}{x - \omega_j}\notag\\
			&+\frac{\eta_5(x)(x - \omega_j)}{x - \omega_j}\label{eq:s_mu_divide2}
		\end{align}
		Combining \eqref{eq:s_mu_divide1} and \eqref{eq:s_mu_divide2}, this leads to
		\begin{equation*}
			T(\omega_j)\prod_{l = n - k}^{\epsilon - 1}(x - \omega_l) - s_{\mu}(\omega_j) - \eta_5(x)(x - \omega_j) = 0.
		\end{equation*}
		Thus one must have $\deg(\eta_5(x)) <  \deg(\prod_{l = n - k}^{\epsilon - 1}(x - \omega_l))$. Hence, according to \eqref{eq:new_divide1}, $\frac{T(x) - T(\omega_j)}{x - \omega_j}$ is the quotient of $\frac{s_{\mu(x)} - s_{\mu(\omega_j)}}{x - \omega_j}$ divided by $\prod_{l = n - k}^{\epsilon - 1}(x - \omega_l)$. By summation on $j$, one can easily verify that the claim holds.
	\end{proof}
\end{lemma}

Lemma \ref{lem:s1x} and \ref{lem:sx} say that $\mathbf{S}(x)$ is the quotient of $\mathbf{S}_1(x)$ divided by $\prod_{l = n - k}^{\epsilon - 1}(x - \omega_l)$, where $\mathbf{S}_1(x)$ is the quotient of $f(x)$ divided by $\bar{\mathbb{X}}_{2^m - 2^{\mu}}(x)$. Therefore, the generalized syndrome $\mathbf{S}(x)$ is determined by the high degree part of $f(x)$. Since $f(x)$ is the IFFT of the vector $\mathbf{r}^{'}$, we have shown that the specific generalized syndrome can be related to the IFFT of $\mathbf{r}^{'}$. Furthermore, as a linear transform relates to any two syndrome definitions, the above comment also proves that any syndrome definition can be characterized in terms of the IFFT regardless of code definition. In the next section, we provide an efficient decoding algorithm base on $\mathbf{S}(x)$.

\section{Proposed Decoding Algorithm}\label{sec:dec}

This section proposes a unified decoding algorithm for $GRS_k(\Ls, \mathbf{w})$ and $\mathcal{A}_k(\Ls, \mathbf{y})$ codes. The term "unified" means that no matter what $\Ls$ and $\mathbf{y}$ are, the algorithm is able to decode the received vector. Furthermore, its computational complexity is $O(n\log(n-k) + (n-k)\log^2(n-k))$ (Note that we assume that $n > 2^{m - 1}$ here. This assumption is reasonable since if $n \leq 2^{m - 1}$, one can construct the code in a smaller field $GF(2^{m-1})$).

Based on the generalized syndrome, the decoding algorithm can be outlined as four steps: 
\begin{enumerate}
	\item computing the generalized syndrome $\mathbf{S}(x)$; 
	\item solving the key equation
	\begin{equation*}
		\mathbf{S}(x)\lambda(x) = q(x)T(x) + z(x);
	\end{equation*}
	\item determining the error locations by finding the roots of $\lambda(x)$; 
	\item computing the error values for the corresponding nonbinary codes.
\end{enumerate}
We provide the detailed algorithm for these four steps in the following. It should be noted that for any positive integer $\upsilon\leq 2^m$, there exist $\upsilon$-points Fourier transform $\text{FFT}_{\bar{\mathbb{X}}}$ and its inverse transform $\text{IFFT}_{\bar{\mathbb{X}}}$, whose complexity are $O(\upsilon\log(\upsilon))$. These transforms are referred to in Appendix of \cite{tang2022a}.

\subsection{Syndrome Computation}
Let $\mathbf{r}^{'}_{l,\epsilon}$ denote a sub-vector of $\mathbf{r}^{'}$:
\begin{equation*}
	\mathbf{r}^{'}_{l,\epsilon} = (r^{'}_{l\cdot\epsilon}, r^{'}_{l\cdot\epsilon + 1}, \dots, r^{'}_{l\cdot\epsilon + \epsilon - 1}).
\end{equation*}
\begin{lemma}
	The coordinate vector of $\mathbf{S}_1(x)$ with respect to $\bar{\mathbb{X}}$ is equal to
	\begin{equation}\label{eq:syndrome_compute}
		\sum_{l = 0}^{2^{m - \mu} - 1}\IFFT_{{\bar{\mathbb{X}}}}(\mathbf{r}^{'}_{l,\epsilon}, \mu, \omega_{\epsilon \cdot l})/p_{2^m - 2^{\mu}}.
	\end{equation}
	\begin{proof}
		Recall that we have $f(\omega_j)  = {r}^{'}_j$ for all $j$. Thus the coordinate vector of $f(x)$ with respect to $\bar{\mathbb{X}}$ can be computed by $\IFFT_{{\bar{\mathbb{X}}}}(\mathbf{r}^{'}, m, \omega_0)$. For any $0\leq \mu \leq m$ and $\epsilon = 2^{\mu}$, the vector
		\begin{equation*}
			(\bar{f}_{2^m - \epsilon}, \bar{f}_{2^m - \epsilon + 1}, \dots, \bar{f}_{2^m - 1}) = \sum_{l = 0}^{2^{m - \mu} - 1}\IFFT_{{\bar{\mathbb{X}}}}(\mathbf{r}^{'}_{l,\epsilon}, \mu, \omega_{\epsilon \cdot l}).
		\end{equation*}
		This property is referred to \cite[Lemma 10]{FFT2016}.
		
		On the other hand, the polynomial $f(x)$ can be represented as
		\begin{align*}
			f(x) = &\sum_{j = 0}^{2^m-1}\bar{f}_j\bar{X}_j(x)\\
			= &\sum_{j = 0}^{2^m-\epsilon-1}\bar{f}_j\bar{X}_j(x) + \sum_{j = 2^m-\epsilon}^{2^m - 1}\bar{f}_j\bar{X}_j(x)\\
			= &\sum_{j = 0}^{2^m-\epsilon-1}\bar{f}_j\bar{X}_j(x) + \bar{X}_{2^m-\epsilon}(x)\sum_{j = 2^m-\epsilon}^{2^m - 1}\bar{f}_j\bar{X}_{j - 2^m-\epsilon}(x)\\
			= &\sum_{j = 0}^{2^m-\epsilon-1}\bar{f}_j\bar{X}_j(x) \\
			&+ {X}_{2^m-\epsilon}(x)\sum_{j = 2^m-\epsilon}^{2^m - 1}\frac{\bar{f}_j}{p_{2^m - 2^{\mu}}}\bar{X}_{j - 2^m-\epsilon}(x)\\
			= &\sum_{j = 0}^{2^m-\epsilon-1}\bar{f}_j\bar{X}_j(x) + {X}_{2^m-\epsilon}(x)\sum_{j = 0}^{2^{\mu} - 1}\frac{\bar{f}_{j + 2^m - 2^{\mu}}}{p_{2^m - 2^{\mu}}}\bar{X}_{j}(x)
		\end{align*}
		According to \eqref{eq:synd1}, we have
		\begin{align*}
			\mathbf{S}_1(x) &= \sum_{j = 0}^{2^{\mu} - 1}\frac{\bar{f}_{j + 2^m - 2^{\mu}}}{p_{2^m - 2^{\mu}}}\bar{X}_{j}(x)
		\end{align*}
		Thus, the coordinate vector of $\mathbf{S}_1(x)$ with respect to $\bar{\mathbb{X}}$ is equal to
		\begin{equation*}
			\sum_{l = 0}^{2^{m - \mu} - 1}\IFFT_{{\bar{\mathbb{X}}}}(\mathbf{r}^{'}_{l,\epsilon}, \mu, \omega_{\epsilon \cdot l})/p_{2^m - 2^{\mu}}.
		\end{equation*}
		This completes the proof.
	\end{proof}
\end{lemma}

Given the coordinate vector of $\mathbf{S}_1(x)$, it remains to calculate $\mathbf{S}(x)$. Recall the equation \eqref{eq:syndrome2}:
\begin{equation*}
	\mathbf{S}_1(x) = \mathbf{S}(x)\prod_{l = n - k}^{\epsilon - 1}(x - \omega_l) +\eta_4(x).
\end{equation*}
One has $\eta_4(\omega_l) = \mathbf{S}_1(\omega_l)$ for $l = n - k, \dots, \epsilon - 1$. As $\deg(\eta_4(x)) < \epsilon - n + k$, $\eta_4(x)$ can be determined by $\epsilon - n + k$ points $\text{IFFT}_{\bar{\mathbb{X}}}$. Furthermore, we have
\begin{equation}\label{eq:syndrome3}
	\mathbf{S}(\omega_j) = (\mathbf{S}_1(\omega_j) - \eta_4(\omega_j))(\prod_{l = n - k}^{\epsilon - 1}(\omega_j - \omega_l))^{-1}.
\end{equation}
Then we can obtain $\mathbf{S}(\omega_j)$ for $j = 0, 1, \dots, n - k - 1$. Finally, the polynomial $\mathbf{S}(x)$ can be computed by $\text{IFFT}_{{\bar{\mathbb{X}}}}$.

We now analyze the computational complexity. Computing the coordinate vector of $\mathbf{S}_1(x)$ costs $O(n\log(n-k))$ operations and evaluating $\mathbf{S}_1(x)$ on $\omega_0, \omega_1, \dots, \omega_{\epsilon - 1}$ costs $O((n-k)\log(n-k))$ operations. Next, determining $\eta_4(x)$ by $\text{IFFT}_{{\bar{\mathbb{X}}}}$ and evaluating it on $\omega_0, \omega_1, \dots, \omega_{\epsilon - 1}$ takes $O((n-k)\log(n-k))$ operations. Finally, the complexity of computing $\mathbf{S}(\omega_j)$ for $j = 0, 1, \dots, n - k - 1$ according to \eqref{eq:syndrome3} is $O(n)$ and the complexity of obtaining $\mathbf{S}(x)$ by $\text{IFFT}_{{\bar{\mathbb{X}}}}$ is $O((n-k)\log(n-k))$. Hence, the complexity of syndrome computation is $O((n-k)\log(n-k))$.

\subsection{The Key Equation}

The Key equation is 
\begin{equation}\label{eq:key_equation}
	\mathbf{S}(x)\lambda(x) = q(x)T(x) + z(x).
\end{equation}
The Euclidean algorithm can solve this key equation. Furthermore, recall that $T(x) = \prod_{j = 0}^{n - k - 1}(x - \omega_j)$. This key equation is an interpolation problem and can be solved by the modular approach within $O((n-k)\log^2(n-k))$ field operations. More details about the modular approach are referred to \cite{tang2022a}.

\subsection{Chien Search}
Once the error locator polynomial $\lambda(x)$ has been found. We can compute the roots of $\lambda(x)$ by
\begin{equation}\label{eq:FFT_search}
	\FFT_{{\bar{\mathbb{X}}}}(\bar{\mathbf{\lambda}}, \mu, \omega_{\epsilon \cdot l}), 0\leq l < 2^{m - \mu},
\end{equation}
where $\bar{\mathbf{\lambda}}$ represents the coordinate vector of $\lambda(x)$ with respect to $\bar{\mathbb{X}}$. The step takes $O(n\log(n-k))$ operations.

\subsection{Forney's formula}

For nonbinary codes, we should compute the error values for each error location. If $\omega_j$ is a root of $\lambda(x)$, the formula for computing the error value at location $\omega_j$ is
\begin{equation}\label{eq:Forney}
	e_j = \frac{q(\omega_j)}{y_i^{-1}\lambda{'}(\omega_j)}.
\end{equation}
Clearly, the computational complexity of this step is at most $O((n-k)\log(n-k))$.

To sum up, the complexity of the proposed unified decoding algorithm is $O(n\log(n-k) + (n-k)\log^2(n-k))$.

\begin{algorithm}[h]
	
	\caption{{Unified Decoding Algorithm}}
	\label{alg:decoding}
	\begin{algorithmic}[1]
		\REQUIRE
		Received vector $\mathbf{r} = \mathbf{c} + \mathbf{e}$.
		\ENSURE
		The codeword $\mathbf{c}$.
		\STATE Compute the syndrome polynomial $\mathbf{S}_1(x)$ according to \eqref{eq:syndrome_compute}.
		\STATE Evaluate $\mathbf{S}_1(x)$ at points $\omega_0, \omega_1, \dots, \omega_{\epsilon-1}$ by Algorithm \ref{FFTX}.
		\STATE Given $\eta_{4}(\omega_{l}) = \mathbf{S}_1(\omega_{l})$ for $l = n - k, \dots, \epsilon - 1$, call $\epsilon - n + k$ points $\text{IFFT}_{\bar{\mathbb{X}}}$ to get $\eta_{4}(x)$.
		\STATE Evaluate $\eta_{4}(x)$ at $\omega_0, \dots, \omega_{\epsilon - 1}$ by $\epsilon$-points FFT and compute $\mathbf{S}(\omega_l)$ for $l = 0, 1, \dots, \epsilon - 1$ according to \eqref{eq:syndrome3};
		\STATE Solve the key equation \eqref{eq:key_equation} by the Euclidean algorithm or the modular approach.
		\STATE Find the error locations by \eqref{eq:FFT_search}.
		\STATE Compute the error pattern $\mathbf{e}$ by \eqref{eq:Forney}.
		\RETURN
		$\mathbf{r} + \mathbf{e}$.
	\end{algorithmic}
	
\end{algorithm}

\section{Separable Goppa Codes}\label{sec:Goppa}

In this section, we discuss the decoding algorithm for the most common separable Goppa codes. Furthermore, it is straightforward to generalize the following discussion for separable Goppa codes to general Goppa codes.

Given $\Ls = (\alpha_0, \alpha_1, \dots, \alpha_{n-1})$, for any vector $\mathbf{a} = (a_0, a_1, \dots, a_{n-1})$ over $GF(2)$, we associate the rational function
\begin{equation*}
	R_{\mathbf{a}}(x) = \sum_{i = 0}^{n - 1}\frac{a_i}{x - \alpha_i}.
\end{equation*}

The Goppa code $\Gamma(\mathcal{L}, G)$ consists of all vectors $\mathbf{a}$ such that
\begin{equation*}
	R_{\mathbf{a}}(x) \equiv 0\bmod{G(x)},
\end{equation*}
in which the polynomial $G(x)\in GF(2^m)$ satisfying $G(\alpha_i)\neq 0$ for $0 \leq i < n$. $G(x)$ is called the Goppa polynomial. If $G(x)$ has no multiple roots, the code is called \textit{separable}. If $G(x)$ is irreducible, the code is called \textit{irreducible}. Evidently, an irreducible Goppa code is separable. 

We assume that $G(x)$ has no multiple roots hereafter, and thus, the corresponding Goppa code is separable.
Given a codeword $\mathbf{a} = (a_0, a_1, \dots, a_{n-1})$ of weight $\kappa$ in $\Gamma(\mathcal{L}, G)$, let $a_{i_0}, a_{i_1}, \dots, a_{i_{\kappa-1}}$ represent the nonzero elements. Define
\begin{equation*}
	\gamma(x) = \prod_{l = 0}^{\kappa - 1}(x - \alpha_{i_l}).
\end{equation*}
The formal derivative of $\gamma(x)$ is
\begin{equation*}
	\gamma^{'}(x) = \sum_{l = 0}^{\kappa - 1}\prod_{j \neq l}(x - \alpha_{i_j}).
\end{equation*}
It follows that
\begin{equation*}
	R_{\mathbf{a}}(x) =  \frac{\gamma^{'}(x)}{\gamma(x)}.
\end{equation*}
As $G(\alpha_i) \neq 0$ for all $i$, $G(x)$ is relatively prime to $\gamma(x)$. Therefore we have $G(x)\mid \gamma^{'}(x)$ since $R_{\mathbf{a}}(x) \equiv 0\bmod{G(x)}$. Note that the coefficients of these polynomials are in a field of characterizing $2$. So there are only even power in $\gamma^{'}(x)$ and $\gamma^{'}(x)$ is a perfect square. Because $G(x)$ has no multiple roots, $\bar{G}(x) = G^2(x)$ is the lowest degree perfect square that is divisible by $G(x)$. It follows that $\bar{G}(x)\mid \gamma^{'}(x)$. Due to $\bar{G}(x)$ must be relatively prime to $\gamma(x)$, we can conclude that
\begin{equation*}
	R_{\mathbf{a}}(x) \equiv 0\bmod{G(x)},
\end{equation*}
if and only if
\begin{equation*}
	R_{\mathbf{a}}(x) \equiv 0\bmod{\bar{G}(x)}.
\end{equation*}
This implies that
\begin{equation*}
	\Gamma(\mathcal{L}, G) = \Gamma(\mathcal{L}, \bar{G}).
\end{equation*}

According to the above discussion, the parity check matrix of $\Gamma(\mathcal{L}, G)$ can be written as
\begin{align}\label{eq:H-Goppa}
	H &= \left(
	\begin{array}{cccc}
		y_0 & y_1 & \dots & y_{n - 1} \\
		y_0\alpha_0 & y_1\alpha_1 & \dots & y_{n - 1}\alpha_{n - 1}\\
		\vdots & \vdots & \dots & \vdots\\
		y_0\alpha_0^{2\rho - 1} & y_1\alpha_1^{2\rho - 1} & \dots & y_{n - 1}\alpha_{n - 1}^{2\rho - 1}
	\end{array}
	\right),
\end{align}
where $y_i = \bar{G}(\alpha_i)^{-1}$ for all $i$ and $\rho = \deg(G(x))$. Hence, the decoding algorithm proposed in Section \ref{sec:dec} can be used for $\Gamma(\mathcal{L}, G)$ and any error with weight less than or equal to $\rho$ can be corrected.

The traditional encoding algorithm for Goppa codes is to extend the $H$ matrix given in ~\eqref{eq:H-Goppa} into a binary matrix $H'$ by replacing each entry in $H$ with its corresponding $m$-tuple over $GF(2)$ arranged in column form. Let $G$ be a binary generator matrix of the Goppa code. Then we have $G(H')^T=\mathbf{0}$, where $T$ is the matrix transpose and $\mathbf{0}$ is the all-zero matrix. Hence, $G$ can be used to encode the information bits.
We next prove that any encoding method resulting in the parity check matrix given in~\eqref{eq:H-Goppa} is suitable for the case when LCH-FFT is being used in the decoding procedure. We prove this fact by giving a systematic encoding procedure to be performed on $\bar{\mathbb{X}}$, which produces the same codeword as the encoding method, resulting in the parity check matrix given in ~\eqref{eq:H-Goppa}. Note that the systematic encoding procedure is an alternative encoding algorithm based on LCH-FFT.

Let $\mathbf{u}=(u_0,u_1,\ldots,u_{k-1})$ be the binary information vector. For ease of presentation, we assume $k$ is a power of $2$. The following is the systematic encoding procedure.
\begin{enumerate}
	\item Calculate 
	$$w_i=\bar G(\alpha_i)\left(\prod_{0\le j\le n-1,j\neq i}(\alpha_i-\alpha_j)\right)^{-1},\ 0\le i\le n-1.$$ This step can be pre-calculated before the decoding procedure.
	\item Calculate $\mathbf{u}'=(u_0',u_1,\ldots, u_{k-1}')$, where
	$$u_i'=u_iw_{\pi(i)}^{-1}.$$
	\item Perform IFFT on $\mathbf{u}'$ to obtain
	$$\bar{\mathbf{u}}=\text{IFFT}_{\bar{\mathbb{X}}}(\mathbf{u}', \log k, \omega_0).$$
	\item Perform FFT on different $\beta$ to obtain
	$$\bar{\mathbf{c}}=(\mathbf{u}',\text{FFT}_{\bar{\mathbb{X}}}(\bar{\mathbf{u}}, \log k, \omega_{k}),\text{FFT}_{\bar{\mathbb{X}}}(\bar{\mathbf{u}}, \log k, \omega_{2k}),\ldots, \text{FFT}_{\bar{\mathbb{X}}}(\bar{\mathbf{u}}, \log k, \omega_{(n/k-1)k}))$$
	when $k|n$; otherwise
	$$\bar{\mathbf{c}}=(\mathbf{u}',\text{FFT}_{\bar{\mathbb{X}}}(\bar{\mathbf{u}}, \log k, \omega_{k}),\text{FFT}_{\bar{\mathbb{X}}}(\bar{\mathbf{u}}, \log k, \omega_{2k}),\ldots, \{\text{FFT}_{\bar{\mathbb{X}}}(\bar{\mathbf{u}}, \log k, \omega_{\lfloor n/k\rfloor k})\}_r),$$
	where the operation $\{\cdot\}_r$ takes the first $r$ element of the vector and $r$ is the reminder when $n$ is divided by $k$.
	\item Multiply each element of $\bar{\mathbf{c}}$ by $w_{\pi(i)}$ to obtain $\mathbf{c}'$, where 
	$$c'_i=\bar{c}_iw_{\pi(i)}.$$
	\item Obtain the codeword $\mathbf{c}$ by permuting elements of $\mathbf{c}'$ by $\pi^{-1}$, where
	$$c_i=c_{\pi^{-1}(i)}.$$
\end{enumerate}
The overall complexity of the encoding procedure is $O(n\log k)$.

Complexity comparisons are provided in Table \ref{tab:Goppa3488}  and Table \ref{tab:Goppa8192}. The code parameters are chosen from the PQC scheme that has been submitted to NIST; see \cite{NIST_PQC} for more details. The code in Table \ref{tab:Goppa3488} is defined over $GF(2^{12}) = GF(2)[x] / (x^{12} + x^3 + 1)$, which is of length $3488$. The Goppa polynomial is $y^{64} + y^3 + y + x$, and its error correction capability is equal to $64$. The code in Table \ref{tab:Goppa8192} is defined over $GF(2^{13}) = GF(2)[x] / (x^{13} + x^4 + x^3 + x + 1)$, which is of length $8192$. The Goppa polynomial is $y^{128} + y^7 + y^2 + y + 1$ and its error correction capability is equal to $128$. The conventional method, the Patterson method, refers to Algorithm 4 in \cite{patterson1975algebraic}. The comparisons are by counting the field operations needed for decoding a codeword, which has been added the most errors that can be corrected. The comparisons given in Table \ref{tab:Goppa3488} and Table \ref{tab:Goppa8192} show that the proposed algorithm is five times and ten times less in terms of computational complexity than the traditional methods, respectively.

\begin{table}[h]
	\caption{Complexity Comparison for decoding Goppa code of length $3488$ and correction ability $64$.}
	\begin{center}
		{
			\begin{tabular}{|c|c|c|c|}
				\hline
				{Decoding method} & {Additions} & {Multiplications}&
				{Inversions}\\
				\hline
				MacWilliams method & 693,857 & 689,889 & 3,617 \\
				\hline
				Patterson method & 475,079 & 466,458 & 189\\
				\hline
				Proposed method & 103,720 & 63,568 & 128 \\
				\hline
			\end{tabular}
		}
	\end{center}
	\label{tab:Goppa3488}
\end{table}

\begin{table}[h]
	\caption{Complexity Comparison for decoding Goppa code of length $8192$ and correction ability $128$.}
	\begin{center}
		{
			\begin{tabular}{|c|c|c|c|}
				\hline
				{Decoding method} & {Additions} & {Multiplications}&
				{Inversions}\\
				\hline
				MacWilliams method & 3,235,840 & 3,219,712 & 8,448 \\
				\hline
				Patterson method & 2,206,791 & 2,171,130 & 381\\
				\hline
				Proposed method & 243,176 & 148,976 & 256 \\
				\hline
			\end{tabular}
		}
	\end{center}
	\label{tab:Goppa8192}
\end{table}

\section{Conclusion}\label{sec:conclusion}
This paper characterizes the generalized syndrome in terms of the IFFT of GRS codes and alternant codes and proposes a fast decoding algorithm of complexity $O(n\log(n-k) + (n-k)\log^2(n-k))$, which is the best complexity to date. This algorithm is suitable for all GRS codes and alternant codes. When decoding practical codes, for example, the Goppa codes used in post-quantum cryptography, a significant improvement is obtained in terms of computational complexity. 

Open issues include whether the Guruswami-Sudan algorithm for RS codes can be accelerated by FFT or not and whether the proposed algorithm can be improved further for irreducible Goppa codes.


\ifCLASSOPTIONcaptionsoff
  \newpage
\fi
\bibliographystyle{IEEEtran}
\bibliography{IEEEabrv,refs_final}

\end{document}